# Forward-Wave Amplification Enhanced Radiation in a 1-THz Harmonic Gyrotron

Zi-Chao Gao, Chao-Hai Du, *Senior Member, IEEE*, Fan-Hong Li, Si-Qi Li, and Pu-Kun Liu, *Senior Member, IEEE*

*Abstract*—Among the most promising terahertz (THz) radiation devices, gyrotrons can generate powerful THz-wave radiation in an open resonant structure. Unfortunately, such an oscillation using high-*Q* axial mode has been theoretically and experimentally demonstrated to suffer from strong ohmic losses. In this paper, a solution to such a challenging problem is to include a narrow belt of lossy section in the interaction circuit to stably constitute the traveling wave interaction (high-order-axial-mode, HOAM), and employ a down-tapered magnetic field to amplify the forward-wave component. A scheme based on the traveling-wave interaction concept is proposed to strengthen electron beam-wave interaction efficiency and simultaneously reduce the ohmic loss in a 1-THz third harmonic gyrotron, which is promising for further advancement of high-power continuous-wave operation.

## I. Introduction

GYROTRONS are capable of generating coherent THz radiation from several watts to megawatts level and there have been significant advancements of gyrotrons from sub-THz to THz range [1-9]. Unfortunately, in some cases, THz gyrotrons still suffer from a large proportion of dissipated power owing to ohmic heating. In the experimental research proposed in Ref. [4], a high electron efficiency (about 10%) is predicted by simulation, whereas the ohmic losses decrease the output efficiency down to 1.3%. For the gyrotron operation, the loss-output ratio is approximately expressed by $Q_d/Q_{ohm}$, where $Q_d$ is the diffraction *Q*-values and $Q_{ohm}$ is the ohmic-loss *Q*-values. Therefore, in previous work dealing with ohmic losses problems, researchers mainly concentrate on how to reduce $Q_d$ or increase $Q_{ohm}$. A natural solution to this problem is to use very high-order modes, and thus a highly oversized cavity with large radius [4, 10, 11] to reduce the $Q_{ohm}$. This solution is widely used in gyrotrons designed for controlled thermonuclear fusion plasma applications [12-14]. However, in such an overmoded cavity, the mode spectrum is dense, thereby complicating the mode competition and making the excitation of operating mode especially tough at the cyclotron harmonic operation [15]. Besides, using various types of cavities, such as sectioned cavities [16-19], cavities with axial irregularities [20, 21], or tapered cavities [22, 23], is demonstrated to be able to effectively reduce the $Q_d$ value.

In this work, we demonstrate that the forward-wave interaction scenario using high-order-axial-modes (HOAMs) provides a general way to reduce the ohmic losses and enhance the electron efficiency of THz gyrotron. Two critical problems are specially considered in our study. First, the excitation of HOAMs is naturally unstable owing to the competition from the first axial mode. Second, owing to the weak coupling between the low-*Q* modes and electrons, the total electron efficiency is generally small. It is found that a short lossy section will effectively stabilize the oscillation of HOAMs, and the magnetic field profile with a down-tapered tail can provide an efficient forward-wave amplification area. Consequently, the final output efficiency is nearly twice as much as that of the high-*Q* open cavity design, and the power dissipated by the ohmic heating only accounts for less than 20% of the total efficiency.

## II. Performance of Different Axial Modes

In our investigation, we consider a 1 THz third harmonic gyrotron operating at TE$_{37}$ mode, and the operating parameters are similar to Ref. [4]. The electrons are assumed to execute large cyclotron orbits, the accelerating voltage is 80 kV, the beam current is 1 A, and the pitch factor of electrons is 1.5. Several eigen axial modes are determined by the resonant structure. Traditionally, HOAMs are frequently used in the gyrotron backward-wave oscillator (gyro-BWO) regime, which provides a possible way for the development of frequency-tunable gyrotron. In fact, in the gyrotron traveling-wave tube (gyro-TWT) regime, there is also a possibility for the excitation of HOAMs. The dispersion diagram of HOAMs in gyro-TWT and -BWO regime is presented in Fig. 1(a).

Frequency-domain self-consistent nonlinear theory is used to predict the possible steady states of a gyrotron circuit. In our consideration, the resistivity of the cavity wall is assumed to equal that of pure copper ($\rho_{cu}$). In this condition, the *Q* values of different axial modes are presented in Table I, and the field profiles are correspondingly shown in Fig. 1(b). For a terahertz circuit, the $Q_d$ of the first-order-axial-mode (FOAM) is comparable with the $Q_{ohm}$ value, while the HOAMs operation decreases the $Q_d$ values by one or two orders of magnitude. As we can see from the field profiles shown in Fig. 1(b), the peak

This work was supported in part by the National Natural Science Foundation of China (Grant Nos. 61531002, NSAF-U1830201, 61861130367, and 61971013), Newton Advanced Fellowship from Royal Society (Grant No. NAF/R1/180121) in the United Kingdom, and the National Key Research and Development Program (Grant No. 2019YFA0210203). *(Correponding authors: Chao-Hai Du; Pu-Kun Liu.)*

The authors are with the Department of Electronics, Peking University, Beijing 100871, China (e-mail: duchaohai@pku.edu.cn; pkliu@pku.edu.cn).

of FOAM locates at the middle of the cavity where the wave is mainly amplified; by contrast, the peaks of second or third axial modes move backward, thereby the ohmic heating mostly occurs at the rear of the cavity [see Fig. 2(b) and (c)]. In this way, the share of ohmic losses ($\eta_{loss}/\eta_e$) greatly decreases. Here, the electron efficiency $\eta_e$ the ohmic loss efficiency $\eta_{loss}$ and the output efficiency $\eta_w$ are defined as $\eta_e=P_e/UI=P_{out}+P_{ohm}/UI$, $\eta_{loss}=P_{ohm}/UI$, $\eta_w=P_{out}/UI$ where $P_e$ is the total power extracted from the electrons, $P_{out}$ is the output power, $P_{ohm}$ is the dissipated power on the cavity wall, $U$ is the accelerating voltage, and $I$ is the beam current. Take the results shown in Fig. 2 for an example, the shares of ohmic losses of the first, second, and third axial modes are 63.1%, 26.0%, and 9.9%, respectively. The final output performances of the first and second axial modes are illustrated in Fig. 3. It is shown that the second-order axial mode (SOAM) operation decreases the $\eta_{loss}/\eta_e$ value down to at most 27%. However, as we can see from the output efficiency ($\eta_w$, solid lines) shown in Fig. 3 (a), despite the decreased losses, the final output efficiency of SOAM is similar with (or slightly exceed) that of FOAM. This is because the electron-wave coupling of the second axial mode is weaker than that of the first axial mode, thereby deteriorating the electron efficiency of SOAM [see the dashed lines shown in Fig. 3 (a)].

TABLE I
Q VALUES OF DIFFERENT AXIAL MODES

| axial modes | $Q_d$ | $Q_{ohm}$ |
| --- | --- | --- |
| first | 16604.0 | 17163.0 |
| second | 5328.5 | 17163.0 |
| third | 972.3 | 17163.0 |

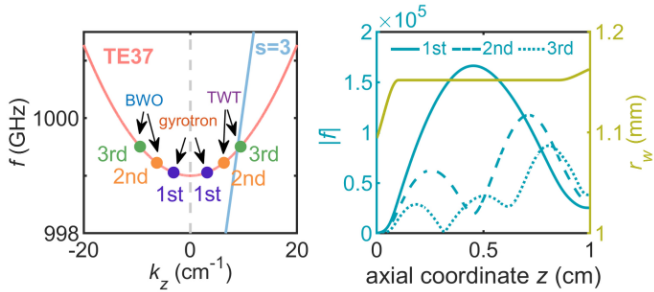

Fig. 1. (a) Dispersion diagram of the gyro-BWO, gyrotron, and gyro-TWT regime. (b) Axial field profile of different axial modes in the gyro-TWT regime.

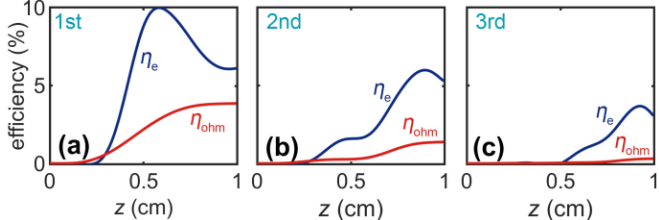

Fig. 2. Electron efficiencies ($\eta_e$) and ohmic losses efficiencies ($\eta_{loss}$) of (a) the first, (b) the second, and (c) the third axial mode. The magnetic field strength is 13.62 T.

### III. AXIAL MODES COMPETITION

In practice, the electrons and the resonate structure will select their favored state and gives out energy, meaning that some of the possible axial modes are unstable in the excitation. Considering that FOAM normally possesses stronger coupling strength with the electrons than HOAMs, the generation of higher modes is generally hard. Here, we adopt a general approach to alleviate the mode competition. The cavity structure with a lossy section is presented in Fig. 4(a). The lossy section whose effective resistivity is $\rho$ provides additional losses only for the FOAM, whereas for HOAMs whose field strength is weak in this section, the oscillation is almost undisturbed. The start oscillation currents (SOC) are shown in Fig. 4(b). By loading a lossy section, the SOC of FOAM increases rapidly in the low-magnetic-field range; by contrast, the SOCs of the second and third axial modes remain nearly undisturbed. As a result, the excitation region of HOAMs is greatly extended.

Time-domain simulation is utilized to further verify the competition situation of SOAM. From the output power shown in Fig. 5(a), for a cavity with homogeneous resistivity, the SOAM will be first excited and then suppressed rapidly by the FOAM. It is likely that the SOAM excited first induces an electron bunching which is also suitable for the FOAM, thereby assisting the excitation of FOAM. When a section with a weak loss [$\rho=500\rho_{cu}$, see Fig. 5(b)] is loaded, the oscillation of SOAM maintains a longer time. When a stronger loss [$\rho=1000\rho_{cu}$, see Fig. 5(c)] is loaded, the SOAM succeeds in the mode competition with FOAM and steadily extracts energy from the electrons. Moreover, although a lossy section is placed in the interaction circuit, there is little increase in the share of ohmic losses of HOAMs [see Fig. 5(d)], because, at the lossy part, both of the propagating and dissipated powers are small.

In the terahertz range, the effective resistivity mentioned in this work can be adjusted by intendedly increasing the material surface roughness [24, 25], or using nanopowder mixture BeO-TiO$_2$-W, where BeO for beryllium oxide, TiO$_2$ for titanium dioxide, and W for tungsten [26]. Two prominent advantages of using a lossy section rather than changing the cavity radius are that it will not introduce extra eigen-axial-modes of the resonant structure and the output power is insensitive to the loss strength (effective resistivity).

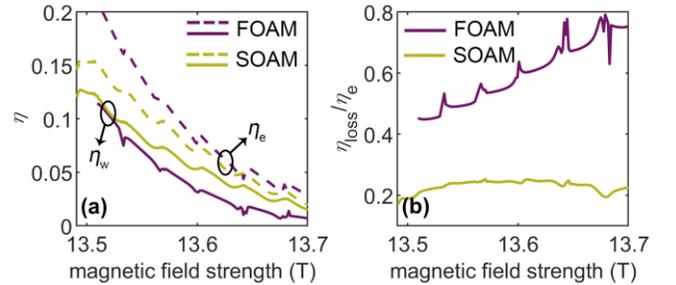

Fig. 3. (a) Output efficiency ($\eta_w$) and electron efficiency ($\eta_e$) and (b) $\eta_{loss}/\eta_e$ values of the first and second axial modes.

### IV. FORWARD WAVE AMPLIFICATION

To compensate for the reduction of electron efficiency induced by HOAMs operation, we should take advantage of the TWT regime operation. A magnetic profile with a tapered tail is adopted to enhance the energy exchange efficiency [27]. The magnetic field profile composed of two parts , i.e. an upstream uniform section and a downstream taper, is shown in Fig. 6(a). The uniform part aims to ensure the excitation of the desired mode and locks the bunching phase of electrons at a certain frequency. After that, the excited wave is amplified through tapered magnetic field profile. From the dispersion diagram shown in Fig. 6, the reduction of magnetic field strength leads to downward translation of the synchronization line, which converts the operation region into the deep gyro-TWT regime. As shown in Fig. 6(a), the electron efficiency is greatly enhanced by the magnetic field profile taper downstream compared with the uniform magnetic field case. In the meanwhile, the ohmic loss efficiency stays nearly unchanged after the nonuniform magnetic field profile is employed.

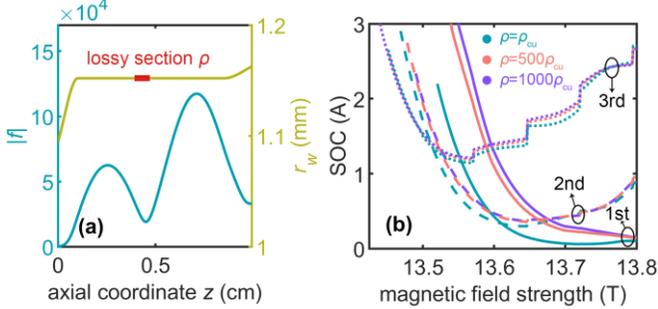

Fig. 4. (a) Structure of cavity with a lossy section whose effective resistivity is $\rho$. (b) Start oscillation current of different axial modes under different $\rho$ values.

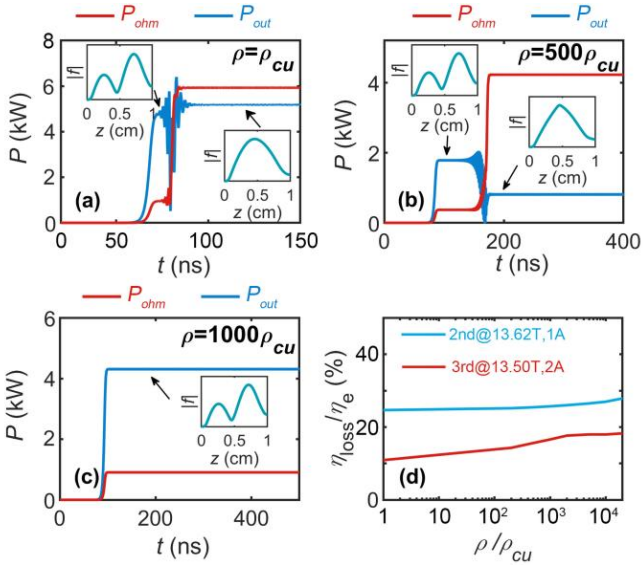

Fig. 5. Output and loss power of the interaction systems shown in Fig. 4 (a) with different effective resistivities. (a) $\rho=\rho_{cu}$. (b) $\rho=500\rho_{cu}$. (c) $\rho=1000\rho_{cu}$. The magnetic field is 13.62 T. The insets indicate the axial field profiles at the time marked by arrows. (d) Share of ohmic losses versus the normalized effective resistivity.

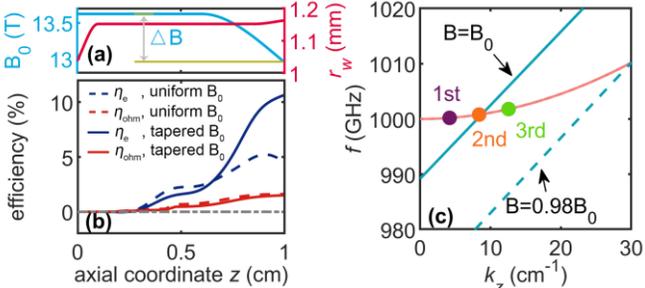

Fig. 6. (a) Cavity configuration with tapered magnetic field profile. (c) The electron efficiencies and ohmic losses of the SOAM under different magnetic field profiles. (c) Dispersion diagram of different magnetic field strengths. The straight lines indicate synchronization lines.

The increment of electron efficiency greatly depends on the gradient of the magnetic field profile, and the results are illustrated in Fig. 7. When the gradient is too large, the electron-wave mismatch will be increased, leading to a deterioration of electron efficiency. Thus, in order to achieve optimized output performance, a proper gradient should be selected. In our investigation, when $\Delta B/B_0$ is around 2%-5%, the output efficiency of the SOAM is nearly twice as much as that of the FOAM [see Fig. 7(a)]. In addition, the utilization of a nonuniform magnetic field can further reduce the ohmic losses [see Fig. 7(b)]. This is because although the wave gains more energy in the tapered magnetic field case, the power flow is compressed to the end of the cavity, thus the dissipated power is nearly the same as the uniform magnetic field case, thereby reducing the $\eta_{loss}/\eta_e$ value.

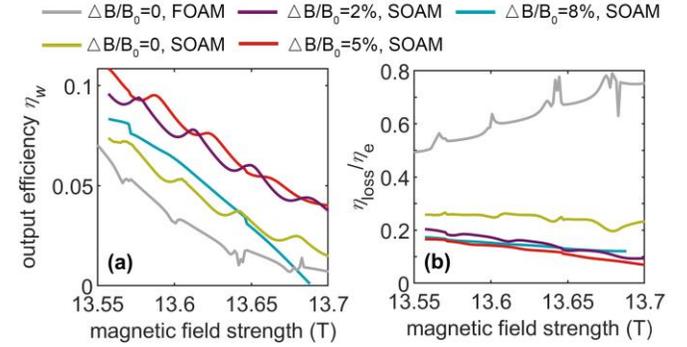

Fig. 7. (a) Output efficiency and (b) share of ohmic losses of the first and second axial modes with different magnetic field gradients.

## V. Conclusion

We propose a high-efficiency terahertz gyrotron scheme with decreased ohmic losses in this work. Unlike the conventional gyrotron operation with a high-$Q$ resonant mode, i.e. FOAM, we attempt to excite low-$Q$ HOAMs to reduce the ohmic loss and constitute a forward-wave amplification interaction. Therefore, we theoretically solve two crucial problems existing in the HOAMs operation. First, the excitation of HOAMs is unstable and encounters competition from FOAM. It is proved that a short lossy section in the front of the cavity will greatly raise the oscillation threshold of FOAM without disturbing the HOAMs. Second, owing to the weakened coupling between the electrons and HOAMs, the electron efficiency is generally lower than that of FOAM. Through a nonuniform magnetic field profile with a down-tapered tail, the forward wave is amplified in the deep TWT regime, thereby enhancing the electron efficiency. In the 1 THz gyrotron operating at the third harmonic, the final output efficiency of the second axial mode is nearly twice as much as that of the first one, and the ohmic loss power only accounts for less than 20% of the total extracted power. The scheme proposed in this work provides a possible way for further developing the low-loss continuous-wave THz gyrotron.